\documentclass[fleqn,usenatbib]{mnras}

\usepackage{newtxtext,newtxmath}

\usepackage[T1]{fontenc}

\DeclareRobustCommand{\VAN}[3]{#2}
\let\VANthebibliography\thebibliography
\def\thebibliography{\DeclareRobustCommand{\VAN}[3]{##3}\VANthebibliography}

\usepackage{graphicx}	
\usepackage{amsmath}	
\usepackage{xcolor}
\usepackage{subcaption}
\usepackage{eso-pic}

\newcommand{\beq}{\begin{equation}}
\newcommand{\eeq}{\end{equation}}
\newcommand{\new}{}
\newcommand{\nSN}{1504 }
\newcommand{\orcid}[1]{\href{https://orcid.org/#1}{\includegraphics[width=8pt]{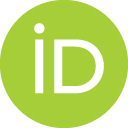}}}

\title[Time Dilation with DES SNe~Ia]{The Dark Energy Survey Supernova Program: Slow supernovae show cosmological time dilation out to $z\sim1$.}

\author[White et al.]{\parbox{\textwidth}{
R.~M.~T.~White,$^{1}$\thanks{E-mail:ryan.white@uq.edu.au}\orcid{0009-0006-7054-0880}
T.~M.~Davis,$^{1}$\orcid{0000-0002-4213-8783}
G.~F.~Lewis,$^{2}$\orcid{0000-0003-3081-9319}
D.~Brout,$^{3}$\orcid{0000-0001-5201-8374}
L.~Galbany,$^{4,5}$\orcid{0000-0002-1296-6887}
K.~Glazebrook,$^{6}$
S.~R.~Hinton,$^{1}$
J.~Lee,$^{7}$\orcid{0000-0001-6633-9793}
C.~Lidman,$^{8,9}$\orcid{0000-0003-1731-0497}
A.~M\"oller,$^{6}$\orcid{0000-0001-8211-8608}
M.~Sako,$^{7}$
D.~Scolnic,$^{10}$\orcid{0000-0002-4934-5849}
M.~Smith,$^{11}$\orcid{0000-0002-3321-1432}
M.~Sullivan,$^{11}$\orcid{0000-0001-9053-4820}
B.~O.~S\'anchez,$^{10,48}$\orcid{0000-0002-8687-0669}
P.~Shah,$^{12}$\orcid{0000-0002-8000-6642}
M.~Vincenzi,$^{13,11}$\orcid{0000-0001-8788-1688}
P.~Wiseman,$^{11}$\orcid{0000-0002-3073-1512}
T.~M.~C.~Abbott,$^{14}$
M.~Aguena,$^{15}$
S.~Allam,$^{16}$\orcid{0000-0002-7069-7857}
F.~Andrade-Oliveira,$^{17,15}$
J.~Asorey,$^{18}$
D.~Bacon,$^{13}$
S.~Bocquet,$^{19}$\orcid{0000-0002-4900-805X}
D.~Brooks,$^{12}$\orcid{0000-0002-8458-5047}
E.~Buckley-Geer,$^{20,16}$\orcid{0000-0002-3304-0733}
D.~L.~Burke,$^{21,22}$
A.~Carnero~Rosell,$^{23,15}$\orcid{0000-0003-3044-5150}
D.~Carollo,$^{24}$
J.~Carretero,$^{25}$\orcid{0000-0002-3130-0204}
L.~N.~da Costa,$^{15}$
M.~E.~S.~Pereira,$^{26}$
J.~De~Vicente,$^{27}$\orcid{0000-0001-8318-6813}
S.~Desai,$^{28}$\orcid{0000-0002-0466-3288}
H.~T.~Diehl,$^{16}$\orcid{0000-0002-8357-7467}
S.~Everett,$^{29}$
I.~Ferrero,$^{30}$
B.~Flaugher,$^{16}$\orcid{0000-0002-2367-5049}
J.~Frieman,$^{16,31}$\orcid{0000-0003-4079-3263}
J.~Garc\'ia-Bellido,$^{32}$\orcid{0000-0002-9370-8360}
E.~Gaztanaga,$^{4,13,5}$\orcid{0000-0001-9632-0815}
G.~Giannini,$^{25,31}$\orcid{0000-0002-3730-1750}
R.~A.~Gruendl,$^{33,34}$
D.~L.~Hollowood,$^{35}$
K.~Honscheid,$^{36,37}$\orcid{0000-0002-6550-2023}
D.~J.~James,$^{53,54}$\orcid{0000-0001-5160-4486}
R.~Kessler,$^{20,31}$\orcid{0000-0003-3221-0419}
K.~Kuehn,$^{38,39}$\orcid{0000-0003-0120-0808}
O.~Lahav,$^{12}$\orcid{0000-0002-1134-9035}
S.~Lee,$^{29}$
M.~Lima,$^{40,15}$
J.~L.~Marshall,$^{41}$\orcid{0000-0003-0710-9474}
J. Mena-Fern{\'a}ndez,$^{42}$\orcid{0000-0001-9497-7266}
R.~Miquel,$^{43,25}$\orcid{0000-0002-6610-4836}
J.~Myles,$^{44}$
R.~C.~Nichol,$^{13}$
R.~L.~C.~Ogando,$^{45}$\orcid{0000-0003-2120-1154}
A.~Palmese,$^{46}$\orcid{0000-0002-6011-0530}
A.~Pieres,$^{15,45}$\orcid{0000-0001-9186-6042}
A.~A.~Plazas~Malag\'on,$^{21,22}$\orcid{0000-0002-2598-0514}
A.~K.~Romer,$^{47}$\orcid{0000-0002-9328-879X}
E.~Sanchez,$^{27}$\orcid{0000-0002-9646-8198}
D.~Sanchez Cid,$^{27}$\orcid{0000-0003-3054-7907}
M.~Schubnell,$^{49}$\orcid{0000-0001-9504-2059}
E.~Suchyta,$^{50}$\orcid{0000-0002-7047-9358}
G.~Tarle,$^{49}$\orcid{0000-0003-1704-0781}
B.~E.~Tucker,$^{9}$
A.~R.~Walker,$^{14}$\orcid{0000-0002-7123-8943}
and N.~Weaverdyck$^{51,52}$
\begin{center} (DES Collaboration) \end{center}
}
\vspace{0.4cm}
\\
\textit{Affiliations are listed at the end of the paper.}
}

\date{Accepted XXX. Received YYY; in original form ZZZ}
\pubyear{2024}

\begin{document}
\label{firstpage}
\pagerange{\pageref{firstpage}--\pageref{lastpage}}
\maketitle

\AddToShipoutPictureBG*{%
  \AtPageUpperLeft{%
    \hspace{0.75\paperwidth}%
    \raisebox{-1.5\baselineskip}{%
      \makebox[0pt][l]{\textnormal{DES-2024-0831}}}
}}%

\AddToShipoutPictureBG*{%
  \AtPageUpperLeft{%
    \hspace{0.75\paperwidth}%
    \raisebox{-2.5\baselineskip}{%
      \makebox[0pt][l]{\textnormal{FERMILAB-PUB-24-0293-PPD}}
}}}%

\begin{abstract}
We present a precise measurement of cosmological time dilation using the light curves of \nSN type Ia supernovae from the Dark Energy Survey spanning a redshift range $0.1\lesssim z\lesssim 1.2$.  We find that the width of supernova light curves is proportional to $(1+z)$, as expected for time dilation due to the expansion of the Universe.  Assuming type Ia supernovae light curves are emitted with a consistent duration $\Delta t_{\rm em}$, and parameterising the observed duration as $\Delta t_{\rm obs}=\Delta t_{\rm em}(1+z)^b$, we fit for the form of time dilation using two methods.  Firstly, we find that a power of $b \approx 1$ minimises the flux scatter in stacked subsamples of light curves across different redshifts. Secondly, we fit each target supernova to a stacked light curve (stacking all supernovae with observed bandpasses matching that of the target light curve) and find $b=1.003\pm0.005$ (stat) $\pm\,0.010$ (sys).  Thanks to the large number of supernovae and large redshift-range of the sample, this analysis gives the most precise measurement of cosmological time dilation to date, ruling out any non-time-dilating cosmological models at very high significance.
\end{abstract}

\begin{keywords}
supernovae: general -- cosmology: observations
\end{keywords}



\section{Introduction}

Time dilation is a fundamental implication of Einstein's theory of relativity in an expanding Universe --- the observed duration of an event, $\Delta t_{\rm obs}$, should be longer than the intrinsic emitted (or rest-frame) duration, $\Delta t_{\rm em}$, by a factor of one plus the observed redshift, $z$,
\beq \Delta t_{\rm obs} = \Delta t_{\rm em}(1+z). \eeq
The idea of using time dilation to test the hypothesis that the Universe is expanding dates back as far as \citet{Wilson1939} and was revisited by \citet{Rust1974}. 
One of the first observational hints of time dilation was the observation by \citet{Piran1992} and \citet{Norris1994} that the duration of gamma-ray bursts (GRBs) was inversely proportional to their brightness -- they used this to argue that at least some GRBs must be cosmological.  The first measurements of cosmological time-dilation using supernovae were made by \citet{Leibundgut1996} for a single Type Ia supernova (SN~Ia) at $z = 0.479$ and \citet{Goldhaber1997} for seven supernovae at $0.3<z<0.5$. Most relevant to this work, we take \citet{Goldhaber2001} as the current state of the art in identifying cosmological time dilation in SN~Ia {\em photometry}. They used 35 supernovae in the redshift range $0.30 \leq z \leq 0.70$, to test a model with a factor $(1 + z)^b$ time dilation and found $b \sim 1.07 \pm 0.06$. 

To avoid degeneracy between the natural variation of light-curve width and time dilation, the evolution of {\em spectral features} of high-$z$ SNe~Ia was used to measure time dilation by \citet{Riess1997AJ}, \citet{Foley2005}, and \citet{Blondin2008}. The first two of these studies found inconsistency with no time dilation at the 96.4\% confidence level (for SN1996bj) and 99\% (for SN1997ex) respectively, and the third finding $b = 0.97 \pm 0.10$. Most recently, \citet{Lewis2023} inferred $b = 1.28^{+0.28}_{-0.29}$ using the variability of 190 quasars out to $z \sim 4$. 
Despite these successes, there remains continued discussion of hybrid or static-universe models such as Tired Light \citep{Zwicky1929,Gupta2023} that do not predict expansion-induced time dilation.  

In this study, we measure cosmological time dilation using SNe~Ia from the full 5-year sample released by the Dark Energy Survey (DES) \citep{DES-SN5YR}, which contains $\sim1500$ SN~Ia spanning the redshift range $0.1 \lesssim z \lesssim 1.2$ --- significantly larger and higher-redshift than any sample of supernovae previously used for a time-dilation measurement.  Such a large sample of SNe is important in reducing statistical uncertainty and such a high-redshift sample is the ideal regime to robustly identify time dilation.  Over half the DES-SN5YR sample are at $z>0.5$, compared to only 12\% in the previous gold-standard Pantheon+ sample \citep{brout19} and 37\% in the Goldhaber analysis \citep{Goldhaber2001}. Therefore the DES sample should have observed light-curve durations more than 1.5 times longer than their rest frame durations (up to 2.2 times longer for those at $z\sim1.2$).  This means their {\em time dilation signal should be significantly larger than the intrinsic width variation} expected due to SNe~Ia diversity in their subtypes.

We test the model that time dilation occurs according to,
\beq \Delta t_{\rm obs} = \Delta t_{\rm em}(1+z)^b. \eeq
If standard time dilation is true we should find $b=1$.  If no time dilation occurs we should find $b=0$.

For this work we aim to keep supernova modelling assumptions to a minimum to avoid circularity in our arguments (because most models of supernova light curves are generated assuming time-dilation occurs).  We therefore take two data-driven approaches to measuring time dilation:
\begin{enumerate}
    \item Firstly, we simply take all the light curves, divide their time axis by $(1+z)^b$ relative to the time at peak brightness, and find the value of $b$ that minimises the flux scatter.
    \item Secondly, we `de-redshift' all the supernova light curves and stack them to define a data-driven SN~Ia `reference light curve'. 
    Then for each individual SN~Ia we measure the observed light curve width ($w$) relative to the appropriate reference, $\Delta t_{\rm obs}=w\Delta t_{\rm reference}$.  This allows us to see if the time-dilation occurs smoothly with redshift and find the best-fit value of $b$, where the expected result would be $w=(1+z)$ corresponding to $b = 1$. 
\end{enumerate}

The first method is entirely data-driven and has no time-dilation assumptions.  In the second method for ease of computation we create the stacked reference by dividing the time axis of the light curves by $(1+z)$.  This method therefore includes an assumption of time-dilation in the generation of the reference.  Even though this assumption is justified by the result of the first method, the second method should strictly be considered a consistency check.  We note it is possible to remove any circularity by keeping the reference light curves in their rest frame and fitting to $w=\left((1+z_{\rm target})/(1+z_{\rm reference})\right)^b$, which requires multiple different reference light curves per target supernova; mathematically this is very similar to our approach but requires even more data.  To further check that this method rules out no time dilation we  re-test method two {\em without} de-redshifting the reference light curves; it dramatically fails the consistency check, see Appendix~\ref{app:nonderedshifted}.     

This paper is arranged as follows.  In Section~\ref{sec:clocks} we discuss the use of type Ia supernovae as standard clocks, and the challenges that need to be taken into account when comparing SNe~Ia light curves observed in different bands across different redshifts. In Section~\ref{sec:data}, we present the data used in this study, while Section~\ref{sec:method} we describe our approach of defining a reference light curve and the determination of the redshift dependence of the time dilation signal. 
We discuss our results in Section~\ref{sec:discussion} and conclude in Section~\ref{sec:conclusions} that the null hypothesis of no time dilation is inconsistent with the data.

\section{Type Ia Supernovae as standard clocks}\label{sec:clocks}
Any investigation into the physics at large length scales in the universe relies on known quantities, be they standard candles, rulers, sirens, or clocks. SNe~Ia have long fit the bill of a standardisable candle on the basis of their extreme brightness and consistency \citep{Tripp1998, Muller-Bravo2022, Scolnic2023}, allowing their observation over cosmic distances with only little uncertainty in their intrinsic properties. As SNe~Ia are the explosions of a white dwarf approaching the Chandrasekhar limit \citep{Hoyle1960, Ruiter2019}, their properties are reasonably uniform across their population compared to other SN types; not only are they standardisable in brightness, but also in time \citep{Phillips1993, Leibundgut1996}. Hence, the observed duration of SN~Ia explosions are well suited to investigating time dilation as a result of an expanding universe \citep{Wilson1939,Rust1974}. 

The presence of a time dilation signal in SNe~Ia data tests the general relativistic prediction of an expanding universe having a factor of $(1 + z)$ time dilation \citep{Wilson1939,Blondin2008}. This signal needs to be corrected for in supernova cosmology analyses \citep{Leibundgut2018,Carr2022} and so conclusively quantifying the effect of time dilation is foundational to our cosmological model, especially considering the continued discussion of hybrid or static-universe models such as Tired Light \citep{Zwicky1929,Gupta2023} that do not predict expansion-induced time dilation.

\subsection{The importance of colour}
\begin{figure}
    \centering
    \includegraphics[width=\columnwidth]{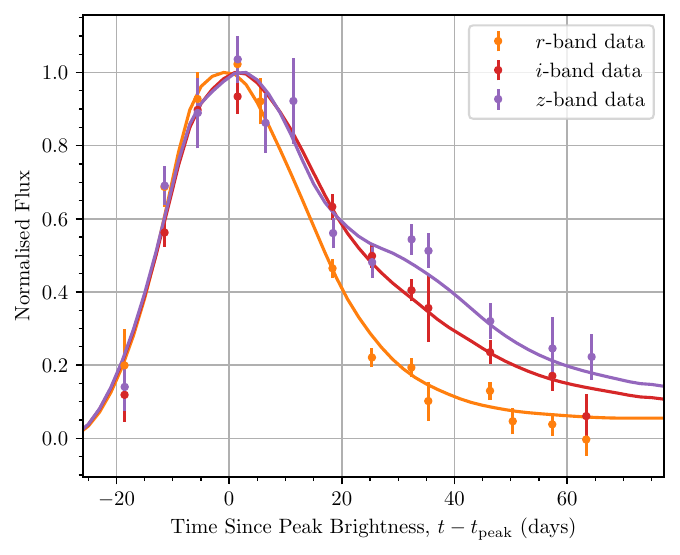}
    \caption{The normalised (in flux) light curve of a SN~Ia at $z=0.4754$ shows intrinsically broader light curves at redder wavelengths that tend to peak later (at least in the optical regime observed across our dataset). The $x$-axis represents time in the observer frame, and SALT3 model fits (solid lines) are overlaid onto the data in each band.}
    \label{fig:single-lightcurve}
\end{figure}
\begin{figure}
    \centering
    \includegraphics[width=\columnwidth]{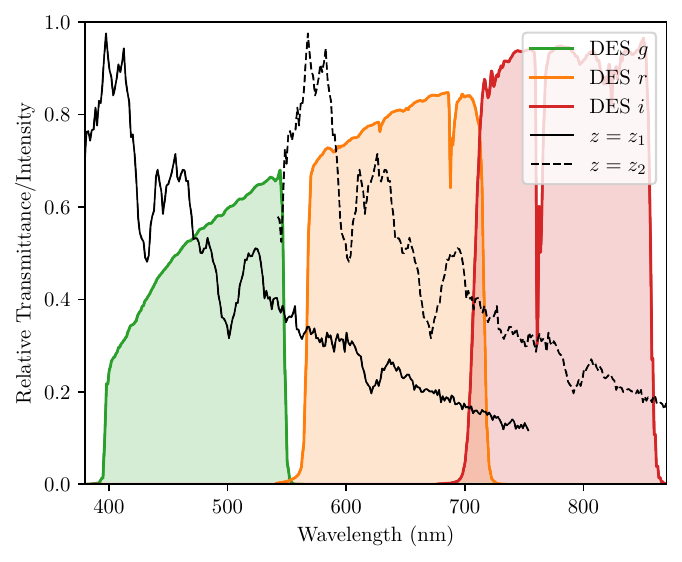}
    \caption{
    For a supernova at redshift $z_1$ observed in a given filter, there exists a higher redshift $z_2 > z_1$ such that another supernova at $z_2$ observed in a redder filter will have a similar rest frame effective wavelength as the nearer supernova.
    We show here the spectrum of SN2001V \citep{Matheson2008} at $z_1\sim0.01$ and again artificially redshifted to $z_2 \sim 0.35$. Here we can clearly see the Si~II absorption line (615nm) redshifted from the $r$ band to roughly the same position along the bandwidth in the $i$ band. Overlaid are the transmission curves of the DES filters from \citet{Abbott2018} Fig. 1. 
    }
    \label{fig:redshift-spectrum}
\end{figure}
SNe~Ia are known to spectrally evolve over the duration of their $\sim 70$ day bright period. 
The early-time spectrum is relatively blue with spectral features dominated by transitions from intermediate mass elements. The spectrum then reddens on the order of days from heavier element emission lines and the cooling of the continuum \citep{Filippenko1997}.
Previous papers have described the redward evolution of SNe~Ia spectra over time \citep{Takanashi2008, Blondin2012, Branch2017}, while photometric evidence of this phenomenon is seen in the light curve peaking later in redder bandpasses than in bluer ones (as in \autoref{fig:single-lightcurve}) for a light curve in the rest-frame optical. As such, the photometric behaviour of a light curve is dependent on the rest-frame wavelength range observed. This means that, to a good approximation, a typical high-$z$ SN~Ia observed in a redder band should have the same photometric and spectral characteristics as a medium-$z$ SN~Ia observed in a bluer band (\autoref{fig:redshift-spectrum}).  Since our photometric bands are fixed, they sample different rest-frame wavelength ranges as the supernovae are redshifted.  Therefore, it is critical to design a method that ensures time dilation measurements compare light curves measured at similar rest-frame wavelengths.\footnote{A note on language: The phrase `rest-frame' wavelengths arises from the usual assumption that redshifts are due to recession velocities.  The fact redshifts occur is not in question here (so it is fine to use $(1+z)$ to calculate matching rest-frame wavelengths, and this contains no time-dilation assumption). The question is whether that redshift arises due to a recession velocity, which would also cause time-dilation.}

\subsection{Stretch-Luminosity relation}\label{sec:stretch}
Our fitting methods are independent of individual supernova light curve models and are based only on the assumption that supernova light curves are all similar. One {\em dissimilarity} between SNe~Ia is the `stretch' in their light curve -- an intrinsic width variation of up to $\sim20$\% that is separate from time dilation and strongly correlated with the peak brightness \citep{Phillips1999}. With this amount of data at such high redshifts we can simply treat this intrinsic variation as noise. 
The stretch variation between SNe~Ia essentially acts as random (but intrinsic) scatter in the obtained widths of light curves. This does not affect the overall trend of light curve width against redshift. 
We therefore make no correction for the stretch-luminosity relation in this work, to maintain maximal model-independence.  

As long as there is a representative sample of the entire population of SNe~Ia at every redshift, this simple analysis should measure time-dilation without bias. However, Malmquist bias can influence the result since brighter supernovae have wider light curves.  If faint supernovae are under-represented at high-redshifts one might expect a slight bias toward a higher inferred time dilation at high-$z$.  Thankfully, the DES data are well-sampled to such high-$z$ that Malmquist bias has minimal impact on our results. \citet{Moller2022MNRAS} showed that the full stretch distribution is well represented in the DES-SN5YR sample out to $z \sim 1.1$ (see also Fig.~\ref{fig:x1data}), meaning that the stretch-luminosity relation should have a negligible effect on our results. 

Previous studies \citep[e.g.][]{Nicolas2021} have also found that the stretch distribution of the SN~Ia population drifts slightly with redshift (getting wider by $\sim3$\% between $0.0<z<1.4$).  Even though we do not see this drift in the DES-SN5YR sample (see Fig.~\ref{fig:x1data}), we quantify the possible impact of this effect on our time dilation measurement in Appendix~\ref{app:x1drift} and find it to be small.

\section{Data} \label{sec:data}
We exclusively use the data of the 1635 type Ia supernovae measured by the Dark Energy Survey Supernova Program \citep{DES-SN5YR}. The DECam instrument on the 4m Blanco telescope at the Cerro Tololo Inter-American Observatory \citep{Flaugher2015AJ} observed most of the photometrically classified SN~Ia candidates in the $g$, $r$, $i$, and $z$ bands according to the criteria in \citet{Smith2020}. The flux is determined by difference imaging \citep{Kessler2015AJ}. High-redshift SNe~Ia typically show negligible flux in the ultraviolet wavelength region, so $g$ and $r$ band light curves are only useful for SNe at $z\lesssim0.4$ and $z\lesssim0.85$ respectively \citep[see][Fig.~2]{DES-SN5YR}. The SALT3 \citep{Kenworthy2021} template fits with a cadence of 2-day time sampling in each band were available for each SN candidate. We use these fits only to estimate the peak flux of each supernova and the associated time of peak flux (given by the $B$-band maximum time in the observer frame) so that we can normalize the light curves in brightness and in time relative to the peak brightness. We otherwise discard the SALT supernova information.

We performed an initial quality cut on the sample of 1635 SNe~Ia, requiring the probability of being a type Ia \texttt{PROBIa} $>0.5$ as classified with \texttt{SuperNNova} \citep{Moller2020MNRAS, Moller2022MNRAS, Vincenzi2024arXiv}. This kept the sample of usable light curves high while removing possible type II supernova contaminants. We removed individual data points from each light curve that had an error in their flux (\texttt{FLUXCALERR}\footnote{\texttt{FLUXCALERR} is the Poisson error on \texttt{FLUXCAL}, which is the variable used for flux in \texttt{SNANA} corresponding to ${\rm mag}= 27.5 - 2.5\log_{10}({\texttt{FLUXCAL}})$.} for the flux value \texttt{FLUXCAL}) greater than 20; this was done to restrict our fitting to the highest quality observations, particularly cutting those with very low signal-to-noise at high redshift (whose observations had comparatively low \texttt{FLUXCAL}).
In the analysis, we did not attempt to fit SNe light curve widths if their light curve had fewer than 5 data points and if their reference curve had fewer than 100 data points (discussed in Section~\ref{sec:method}). This was done on a per-band basis; we estimated the width of each SN light curve in each band where it satisfied these criteria.
Individual light curves were also omitted from the analysis if the $\chi^2$ width fitting did not converge. All together, after these quality cuts we were left with width measurements of \nSN unique SN~Ia across the dataset.

\section{Fitting Supernova Photometry to a Reference Light Curve} \label{sec:method}
The photometric analysis in \citet{Goldhaber2001} relied fundamentally on template fitting to measure the expansion-induced time dilation signal in SNe~Ia. With the wealth of data now available we can in principle obtain the same dilation signal using the data alone, independent of a light-curve template. Herein we describe such a template-independent method involving scaling the time axis of each of the 1504 individual light curves to fit their own \emph{unique} reference curve composed of all the matching photometry in the DES dataset.  The time dilation signal in each light curve is the multiplicative inverse of said scaling factor, and this can be found for each of the \nSN SN~Ia light curves in the data. This differs from a traditional template-fitting method in that we do not assume the shape of a light curve but instead let the data from other SNe compose the template-like reference curve.  

\subsection{Reference curve construction} \label{sec:ref_curve}
\begin{figure}
    \centering
    \includegraphics[width=\columnwidth]{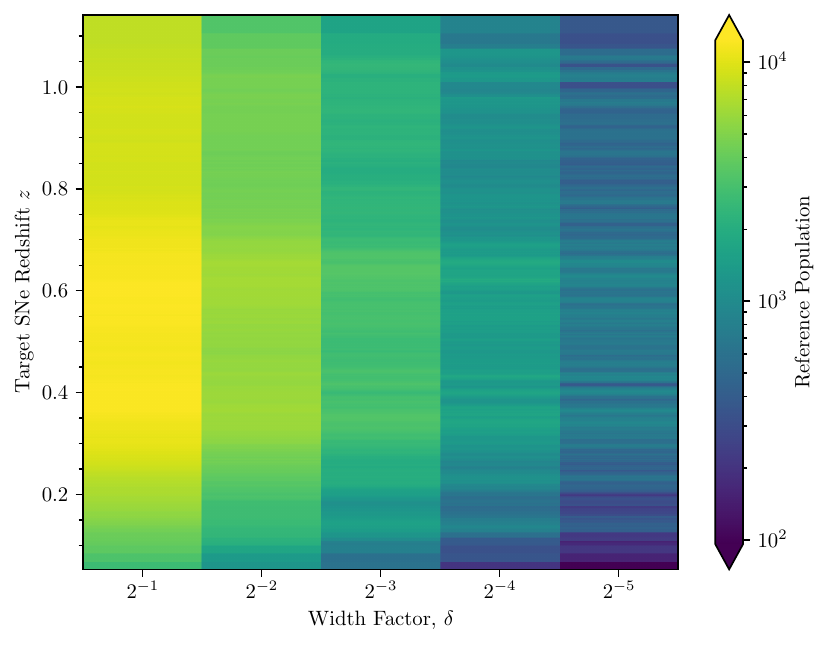}
    \caption{For each of the SNe~Ia in our sample, we constructed a reference light curve with a $\delta = 2^{-x}$ parameter according to \autoref{eq:referencerange}, with $\Delta \lambda_f$ band FWHM. We counted how many data points populated the reference curves (i.e. the number of \emph{points} in Fig.~\ref{fig:referenceconstruction} for example) changing $\delta$ in integer steps of powers of 2. This plot shows the reference populations for a target SN measured in the observer frame $i$ band, and is largely similar for the different bands differing mainly by a linear shift on the vertical axis. That is, an analogous plot in a bluer band would see the colour distribution shifted downwards in target redshift and a redder band upwards.}
    \label{fig:referencepop}
\end{figure}
The main functionality of this method is to use the photometric data of many SNe to automatically create a reference light curve unique to any one target SN. The target SN is the SN whose width we are trying to measure. 

Since the shape of a SN~Ia light curve is dependent on the rest-frame effective wavelength at which it is observed \citep[Fig.~\ref{fig:single-lightcurve}; see also][]{Takanashi2008,Blondin2012}, the reference photometry \emph{must} be composed only of light curves that have the same (or very similar) intrinsic shape as the target SN. Hence, we must choose reference photometry that samples the same rest-frame effective wavelength as the target light curve. This effect is shown in Fig.~\ref{fig:redshift-spectrum}, where, for example, we might compare a low-$z$ supernova in some band to a higher-$z$ supernova observed in a redder band with the same (or similar) rest frame effective wavelength. We can compare the photometry between the two events provided that their rest-frame effective wavelength (and hence their light curve shape/evolution) is alike.

If we chose instead to fit each target light curve in some band against all of the photometry from that band, we would expect a non-linear change in slope as a function of redshift on a width-vs-redshift plot. The explanation for this lies in the fact that SNe~Ia spectra get redder over time; the light curves measured in a redder band are intrinsically wider than those measured in a bluer band as shown in Fig.~\ref{fig:single-lightcurve}. Hence, with this hypothetical method (comparing to all photometry), we would observe a bluer than average rest-frame curve for a high redshift SNe which would bias the obtained width to an intrinsically thinner value. Conversely, we would be biased towards a wider (redder) than average width for low redshift supernovae. To avoid this bias, we use the aforementioned method of only using reference photometry with a similar rest-frame wavelength as our target light curve.

To find relevant light curves to populate the reference curve, we pick all light curves out of a calculated redshift range. To fit a single (target) SN light curve at redshift $z$ imaged in a band of central wavelength $\lambda_f$, we can populate the reference curve with SNe within the redshift range
\begin{equation} \label{eq:referencerange}
    \frac{\lambda_r (1 + z)}{\lambda_f} - \delta\frac{\Delta \lambda_f}{\lambda_f}\leq 1 + z_r \leq \frac{\lambda_r (1 + z)}{\lambda_f} + \delta\frac{\Delta \lambda_f}{\lambda_f}
\end{equation}
whose photometry is measured in a band of central wavelength $\lambda_r$. Here $\delta$ is a free parameter which, together with the band full width at half maximum (FWHM) $\Delta \lambda_f$, describe the acceptable wiggle room in the relative band overlap. A derivation of this formula is given in Appendix~\ref{app:reference} and a graphical representation of this construction is shown on the left-side plots in Fig.~\ref{fig:widthfitting}. 

We show in Fig.~\ref{fig:referencepop} the number of points in the reference curves (hereafter referred to as reference population) for all of the DES SNe with a variable $\delta$ parameter. Ideally, this $\delta$ parameter should be as small as practical to ensure that the reference curve is consistent in shape (i.e. the spread of rest frame effective wavelengths is small). In practice, we find a value of $\delta = 2^{-4}$ is the minimal value that provides a large enough reference population for high/low redshift target SNe (on the order of $\sim 10^2$ needed to satisfy the Section~\ref{sec:data} criteria at $z \sim 1.1$). For medium range target SNe redshifts (in the context of the DES-SN sample), we note that the reference population is large ($\sim10^3$) even for $\delta \lesssim 2^{-4}$. 

After populating the reference curve with data points we then normalise the photometry in flux; as the curve is populated with the data of several SNe at different redshifts, the curve must be homogeneous in flux. To do this, we utilised the peak flux in the SALT3 model light curves provided for each SNe, which is independent of the time dilation signal. The data in each constituent curve is normalised by this value before being added to the reference. For convenience we also use the time of peak brightness given by SALT3 as the reference point about which to stretch the light curves (see equation~\ref{eq:lightcurve}), and this parameter is also in the observer frame of reference. These are the only uses of SALT3 information and we expect that the same time dilation signal would be obtained in the data with any other consistent normalisation measures.

\subsection{First measure of time dilation: minimising scatter in the reference curve} \label{sec:min_scatter}
\begin{figure}
    \centering
    \includegraphics[width=\columnwidth]{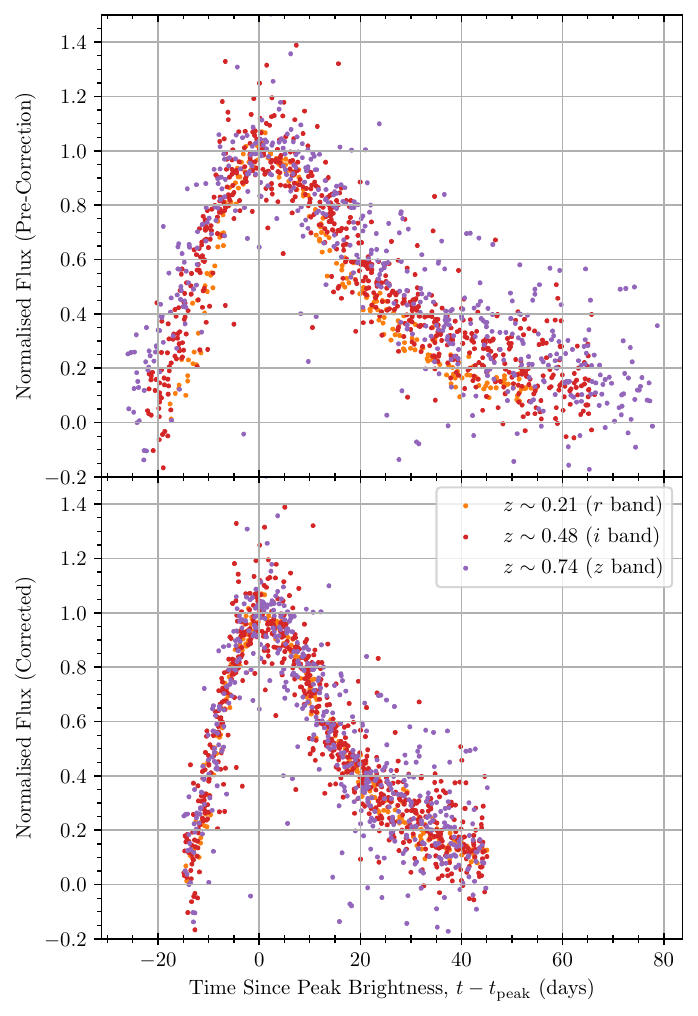}
    \caption{For a target SN~Ia at $z \simeq 0.48$, the $i$-band reference curve consists of data from the $r$, $i$, and $z$ bands. These data are chosen at redshifts according to equation~(\ref{eq:referencerange}) (and visually shown in the left plots of Fig.~\ref{fig:widthfitting}). \emph{Top:} the data in different bands are not in phase (in the observer frame). It is visually obvious that light curves appear wider in time at higher redshift. \emph{Bottom:} after a $(1 + z)$ correction to the SN~Ia light curves, we see a consistent trend across all bandpasses and time. This alone is evidence for some degree of time-dilation.
    }
    \label{fig:referenceconstruction}
\end{figure}
\begin{figure}
    \centering
    \includegraphics[width=\columnwidth]{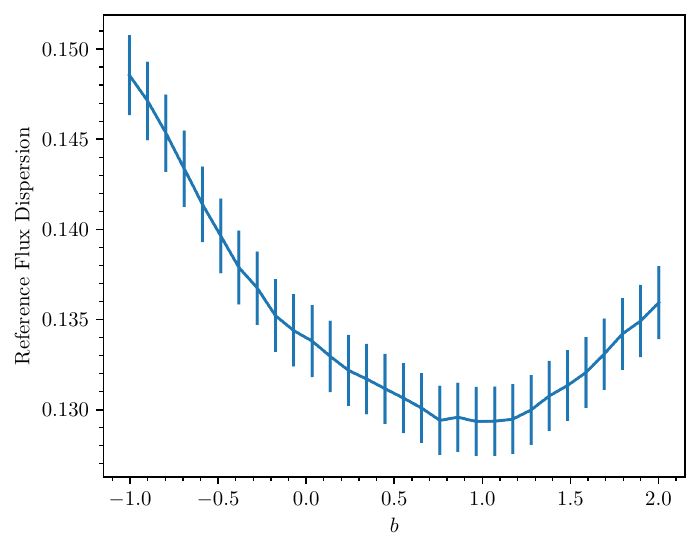}
    \caption{By scaling the reference photometry in time according to $(1 + z)^b$ for some free parameter $b$, we find $b\sim 1$ minimises the reference flux dispersion across the entire SNe sample. The reference flux dispersion represents the median dispersion of flux across the entire sample of normalised reference light curves in each band (here averaged for the $riz$ bands), where the errorbars indicate one standard deviation in these values. We note that this figure yields a signal of $(1 + z)$ time dilation in the DES dataset, independent of the rest of the analysis.}
    \label{fig:PowerDispersion}
\end{figure}

After the flux of the reference curve is normalised, we see that the different bandpass data in the curve are temporally stretched (see the colour gradient of the top plot in Fig.~\ref{fig:referenceconstruction}). As the redder bandpasses are sampled at higher redshift, this is an immediate indication of time dilation. Without assuming our expected cosmological time dilation of $(1 + z)$, we can scale the data in \emph{all} of the reference curves by a factor of $(1 + z_j)^b$, where $z_j$ is the redshift of each constituent curve in a reference and $b$ is a free parameter. We posit that minimising the flux dispersion in the reference curve is analogous to finding the optimal temporal scaling, simultaneously minimising the dispersion in time. Hence, finding the value of $b$ that minimises the flux scatter gives us our correction factor. 

To investigate this, we generated reference curves for each of the target SNe per observing band and scaled the data in time according to the aforementioned relation in terms of $b$ (Fig.~\ref{fig:referenceconstruction} shows this scaling for $b=1$ as an example). Then, we binned the timeseries data into 30 equal-width time bins and found the standard deviation of the flux within each bin. We calculated the median of these 30 standard deviations as a representative estimate of the total flux scatter for that reference curve with that tested $b$ value. We then took the median of those results across all the SN reference curves as our estimate of the dispersion for that $b$, which is shown in Fig.~\ref{fig:PowerDispersion}.
That is, our reference flux dispersion is 
\begin{equation}
    \sigma_\text{rf}(b) = \text{Med}\left(\{\{\text{Med}(\sigma_{ij}(b)) | \, \forall j \in (1, ..., 30) \} | \, \forall i \in (1, ..., N_\text{sn}) \}\right) \label{eq:reffluxdisp}
\end{equation}
for $N_\text{sn}$ supernova light curves in that band, and $\sigma_{ij}(b)$ being the flux standard deviation of the $i$th light curve in the $j$th timeseries bin. This process was repeated for each of the $riz$ observing bands, omitting the $g$ band due to the smaller number of SNe. 
We crudely estimate the error (for each $b$) in this method as being the standard deviation of the median dispersions across all light curves.
We find an optimal scaling corresponding to $b \sim 1$ (Fig.~\ref{fig:PowerDispersion}) across the entire dataset, which is the expected dilation factor of $\sim (1+z)$. 

If there was no time dilation we would expect the minimum dispersion in the reference curve to be at $b\sim0$ (i.e.\ no time scaling) in Fig.~\ref{fig:PowerDispersion}. The fact that we find $b\approx1$ is evidence for time dilation of the expected form.
This \textcolor{red}{is} a rough but completely model independent measure of time dilation and it is the paper's first main result. 

\subsection{Second measure of time dilation: Finding each light curve width} \label{sec:lightcurvewidth}
After constructing the reference curves for a target SN, we are ready to fit for the width, $w$, of each individual target light curve and look for a trend with redshift. This method enables a more precise measure of $b$. 
\begin{figure*}
    \centering
    \includegraphics[width=.74\textwidth]{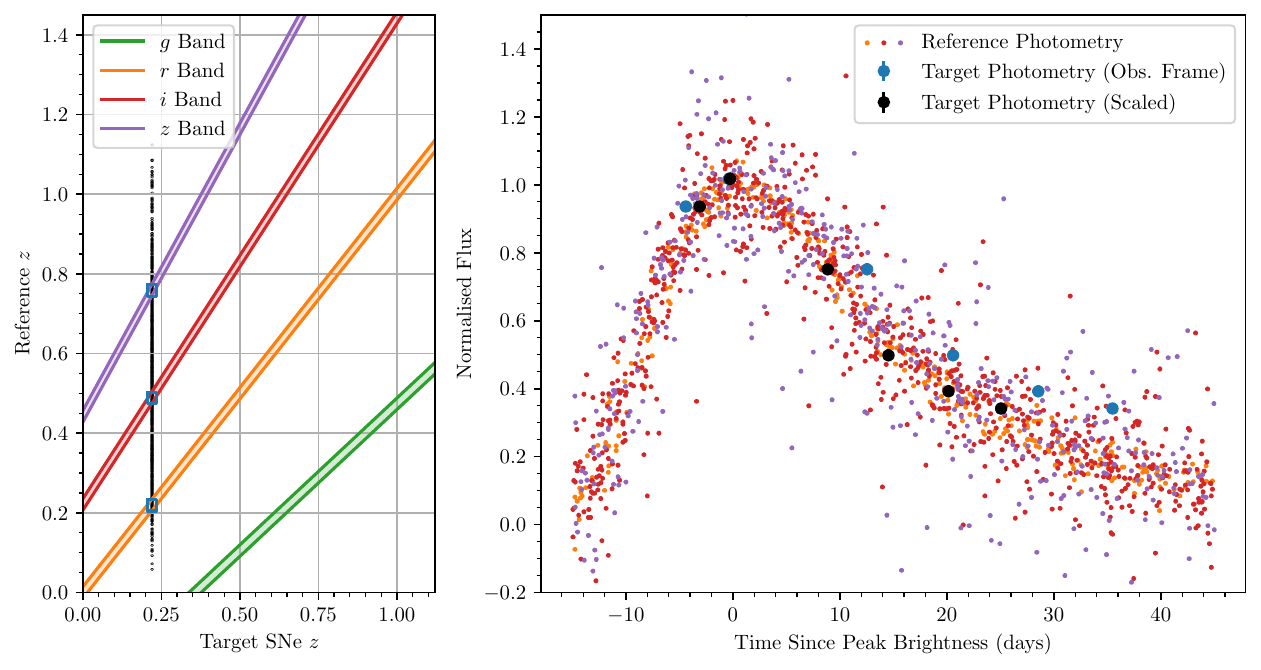}
    \includegraphics[width=.74\textwidth]{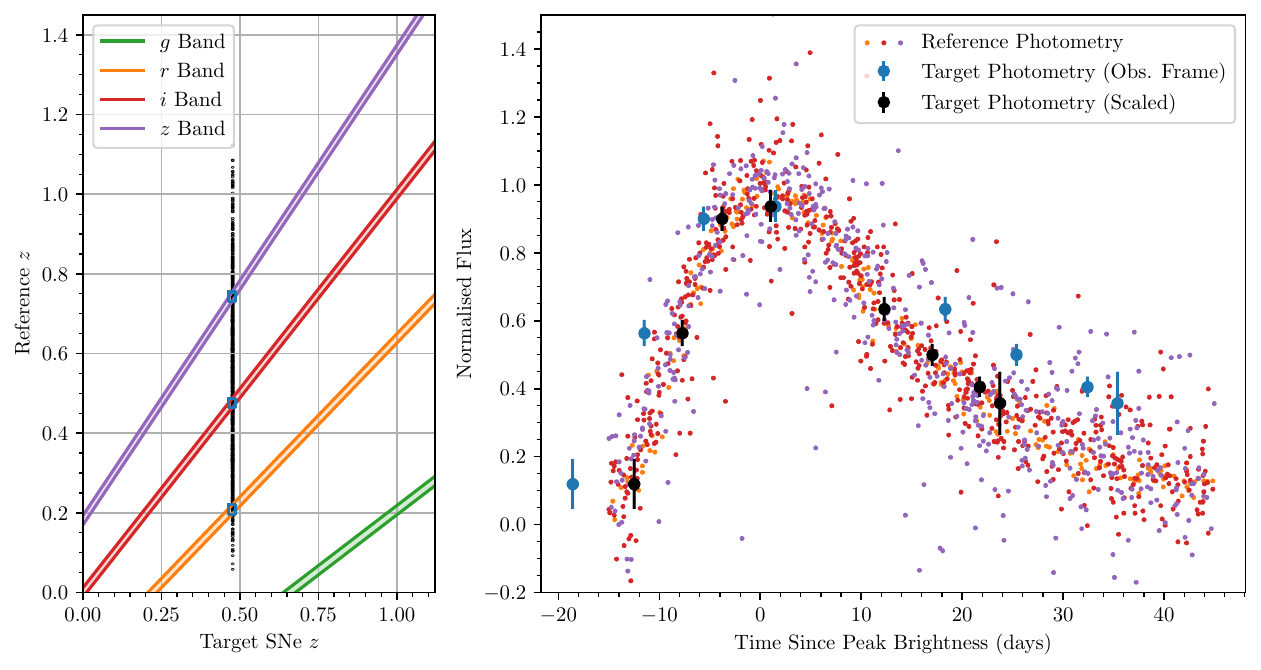}
    \includegraphics[width=.74\textwidth]{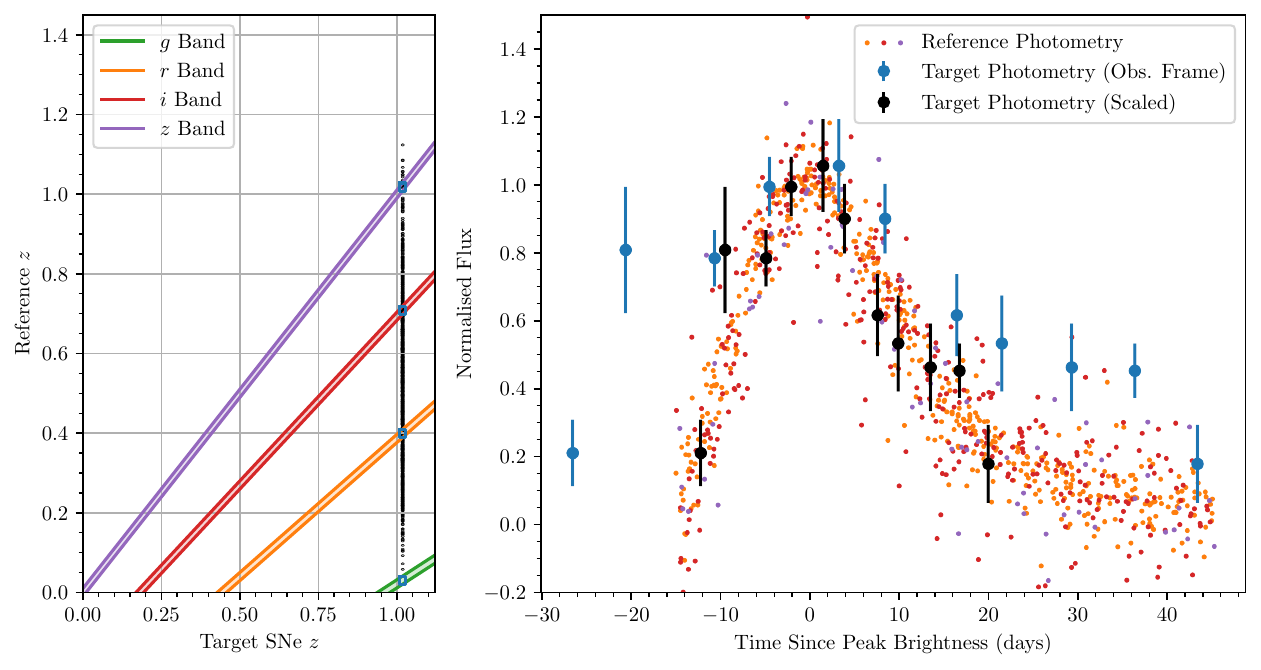}
    \caption{We show the reference curve construction and subsequent target SN fit for 3 SNe at redshifts $z \simeq 0.22$, $z \simeq 0.43$, and $z \simeq 1.02$ and in fitting bands $r$, $i$, and $z$ respectively (in descending order). The left plots show the allowed ranges for reference curve SN sampling given the target redshift (and $\delta = 2^{-4}$). The vertical line of dots is plotted at the target SN redshift, with each dot representing the redshift of a DES supernova (vertical axis).  The dots that fall in the narrow coloured bands are the SNe that make up the reference population, as those data all share approximately the same rest-frame wavelength in their respective bands; boxes are plotted surrounding these regions to more clearly indicate the light curve colour samples that make up the reference curve.  The right plots show the constructed $(1 + z)$ time-scaled reference curve (small coloured points) with respect to the target SN photometry (blue points) and subsequent target photometry scaled on the time axis to fit the reference (best-fit widths of 1.42, 1.49, and 2.17 respectively). Due to the statistics associated with such large reference curve populations, the contribution of any individual reference point uncertainty to the overall reference curve uncertainty is negligible and not plotted; the uncertainty in the target data has a much higher contribution to the uncertainty in the fitting. }
    \label{fig:widthfitting}
\end{figure*}
\begin{figure*}
    \centering
    \includegraphics[width=0.85\textwidth]{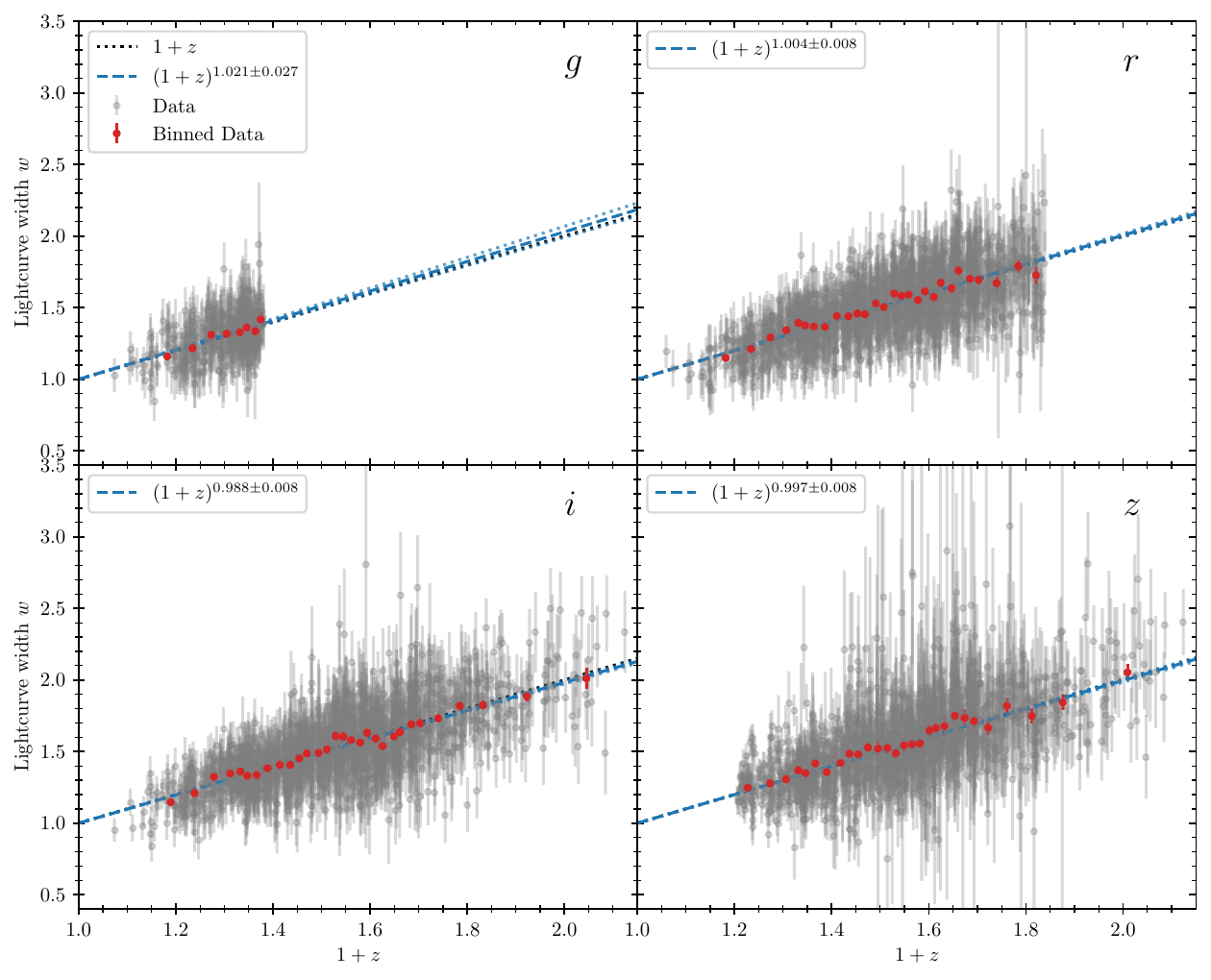}
    \caption{Using the reference-scaling method described in Section~\ref{sec:lightcurvewidth}, we plot the fitted SNe widths of light curves observed in the $g$, $r$, $i$, and $z$ bands (\emph{left} to \emph{right}, \emph{top down} respectively). The lines of best fit (blue dashed) are in excellent agreement with the expected $(1+z)$ time dilation (black dotted). The binned data are purely to visualise rough trends in 50 data point bins. 361 SNe in the $g$ band passed the quality cuts described in Section~\ref{sec:data}, while the $r$ band has 1380 SNe, the $i$ band 1465, and the $z$ band 1381. The reduced chi-square values, $\chi_\nu^2$, of each fit (left to right, top down) are 0.537, 0.729, 0.788 and 0.896 respectively.}
    \label{fig:allwidths}
\end{figure*}
\begin{figure*}
    \centering
    \includegraphics[width=.85\textwidth]{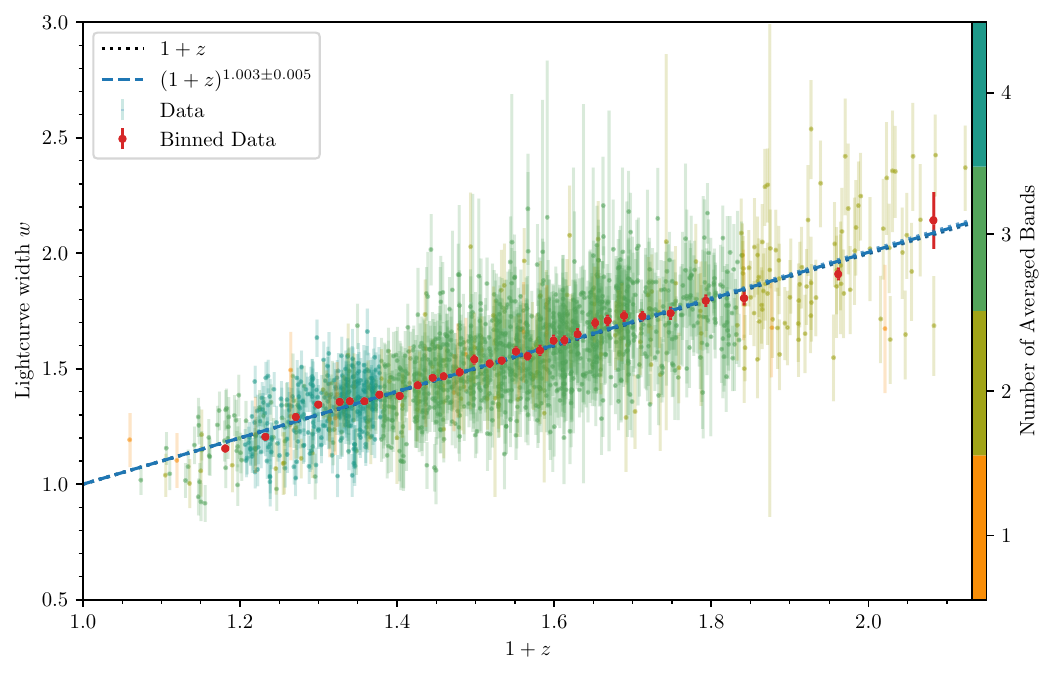}
    \caption{We show here the width value for each SNe averaged across all available bands. Since cosmological time dilation is independent of the observed band of any SN, we are able to average the widths over observed bands to form a more robust estimate of the light curve width. This relationship of $(1 + z)^{1.003\pm0.005}$ time dilation (reduced chi-square $\chi_\nu^2 \simeq 1.441$) is comprised of the \nSN unique SNe across the 4 bandpasses, where the error bars here are the Gaussian propagation of the errors in each band. Points are coloured according to how many bandpasses were used in computing the averaged width. A linear model fit to the data recovers $w = (0.988\pm0.016)(1 + z) + (0.020\pm0.024)$ (with the same $\chi^2_\nu$ to 4 significant figures), consistent with our power model fit above.}
    \label{fig:widthAve}
\end{figure*}

We first normalise the target data to the peak flux using the SALT3 fit (as with the reference curves). The free parameter in the fit is the scaling parameter $1/w$, whereby changing this value would stretch and squash the data relative to $t_\text{peak}$ (the time since peak flux) until the $\chi^2$ is minimised. That is, we assume the SN~Ia light curve of the $i$th supernova is of a mathematical form similar to that described in \citet{Goldhaber2001},
\begin{equation} \label{eq:lightcurve}
    F_i(t) \simeq f_i\left(\frac{t - t_\text{peak}}{w}\right)
\end{equation}
and change $w$ until the data most closely matches the reference. Here, $f_i(t)$ corresponds to the $i$th target light curve; $F_i(t)$ corresponds to the $i$th reference curve where each point is now scaled in time by $(1 + z)$ relative to $t_\text{peak}$ as per the results of Section~\ref{sec:min_scatter}. 

To fit the target light curve width using its reference curve, we minimised the $\chi^2$ value of the differences in the target flux compared to the median reference flux in a narrow bin around time values of the target photometry. That is, for each target light curve we minimised
\begin{equation}
    \chi_i^2 = \sum_j^{N_p} \frac{\left(f_{ij} - \text{Med}\left\{F_i(t)~|~\forall t \in [t_{ij}/w - \tau, \, t_{ij}/w + \tau]\right\}\right)^2}{\sigma_{ij}^2} \label{eq:chisquaremin}
\end{equation}
for $N_p$ number of points in the $i$th target SN light curve ($f_i$). The points in the reference curve ($F_i$) bin that are averaged and compared to each target SN flux value ($f_{ij}$ -- with error $\sigma_{ij}$) are selected within the time range $[t_{ij}/w - \tau, t_{ij}/w + \tau]$; here $t_{ij}$ is the time since peak brightness of each target data point scaled by the fitted width $w$, and $\tau$ is the half bin width either side of the central time value $t_{ij}$. 

During the fitting process, the bounds of this narrow bin around each time value changes as the target data is scaled in time but remains the same width. 
We chose a bin width, $2\tau$, of 4 rest-frame days (i.e. $\pm\tau=\pm 2$ of a central value); ideally this would be as low as practical to maximise intrinsic similarity between the target data point position and the reference curve slice, but needs to be large enough to provide a sufficiently populated sample of the reference to compare to. We find that a width of 4 days (just under the width of a minor tick span in Fig.~\ref{fig:referenceconstruction}) is low enough that the reference curve does not significantly change in flux but still contains enough points even for high/low redshift target SNe with small reference populations. 
With this $\tau = 2$ value we find $\gtrsim 50$ data points per time slice at the highest and lowest redshifts, where a $\tau = 1$ yields a prohibitively small $\lesssim 20$ data points per slice even in the most well sampled photometric band ($i$-band). 

In fitting the data, we did not include any target SN data points that extended past the maximum time value in the reference curve; the late-time light curves of SNe dwindle slowly and are less constraining for width-measurements than those near the peak. We also omitted any points that had observation times prior to the first reference curve point from the fitting procedure. 
 
We note that this method of fitting is not fundamentally limited to target SN data with pre-peak brightness observations in each band. Given enough target data (on the order of several well-spaced points in time), the mapping of this data to the corresponding reference curve phases is unique regardless of whether pre-peak data is available.

The uncertainty in each estimated width was found via Monte Carlo uncertainty propagation, where the target data was resampled 200 times according to its Gaussian error; for each iteration we fit the width and the final error is the standard deviation in these widths. To provide some measure of the error in the reference curve, we imposed an error floor of $\sigma_m(1 + z)$ on the Monte Carlo uncertainty, for $\sigma_m$ representing the median (normalised) flux dispersion in the reference curve. The $\sigma_m$ dependence is from our rationale that the fit width is only as good as the quality of the reference curve, and the $(1 + z)$ dependence arises from our scaling of the reference curve. 

An example of how a reference curve is created is shown in Fig.~\ref{fig:referenceconstruction}. We note the difference in the reference curve timescale before and after $(1 + z)$ correction, and the larger dispersion in flux between the points at any one time prior to correction. Several examples of width fitting and reference curve construction for low, medium, and high redshift target SNe (in the context of the DES-SN sample) are shown in Fig.~\ref{fig:widthfitting}. 

We note that while the width fitting for the whole dataset was calculated in all four DECam bands, only the $i$ band data encompasses the entire redshift range of the DES-SN sample. Due to the spectral shifting inherent in redshifted data, the $g$ and $r$ filters are unable to detect SNe at sufficiently high redshift ($z \gtrsim 0.4$ and $z \gtrsim 0.85$ respectively) as the observed wavelengths shift to lower emitted wavelengths \citep[see Fig. 2 of][]{DES-SN5YR} and become fainter as a result. Fig.~\ref{fig:widthfitting} shows that fitting $z\lesssim0.2$ SNe in the $z$-band would require negative redshifted SNe in the other bands to populate the reference; hence there is an inherent redshift floor for $z$-band fits leaving the $i$-band as the only suitable bandpass for the entire redshift range.

The widths obtained in all four bands separately are shown in Fig.~\ref{fig:allwidths}. We see the truncated $g$, $r$ and $z$ band data, and fit widths consistent with the expected $(1 + z)$ relationship across all bands. The averaged widths of all the bands are shown in Fig.~\ref{fig:widthAve}, again showing excellent agreement with the $(1 + z)$ expected theory.

As mentioned in the introduction, this method has an element of circularity because we de-time-dilated the observed light curves to generate each reference light curve.  As a cross-check to ensure that we are not just getting the answer we put in we repeated the analysis {\em without} de-redshifting the data.  This effectively makes the reference light curves more noisy and wider (like the top plot of Fig.~\ref{fig:referenceconstruction}). If time dilation is absent we should get a consistent $b=0$ fit in this case.  However, if $(1+z)$ time dilation is present we should find a slope {\em in}consistent with $b=0$ and an intercept offset from $w=1.0$ (because the reference light curve will itself be time-dilated). We find, as expected, that the $b=0$ result is excluded strongly by this test (see Appendix~\ref{app:nonderedshifted} and Fig.~\ref{fig:allwidths_nonderedshifted}).

\section{Discussion}\label{sec:discussion}

\begin{figure}
    \centering
    \includegraphics[width=.8\columnwidth]{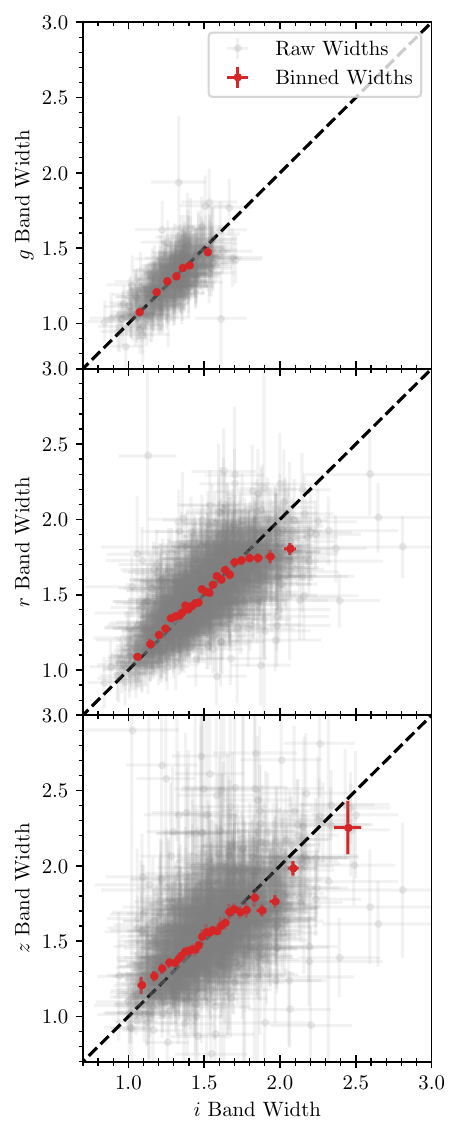}
    \caption{Since most SNe were observed in multiple bands, the fit widths in each band for each SNe should be intrinsically correlated as they arise from the same event. Hence, the widths for the same target SN should show some agreement between bands. We plot their agreement relative to the $i$ band which had the most SNe pass the quality cuts. Each of the data points here corresponds to a width in Fig.~\ref{fig:allwidths} of bands $g$, $r$, or $z$ against the $i$ band widths. A 1:1 dashed line is shown to represent perfect agreement and binned points are plotted to represent the trends in the agreement. }
    \label{fig:WidthComparison}
\end{figure}

As we see in Fig.~\ref{fig:allwidths}, there is a clear and significant non-zero time dilation signature in the DES SN~Ia dataset, conclusively ruling out any static universe models. Our method described in Section~\ref{sec:method} detects a time dilation signature in all of the $g$, $r$, $i$, and $z$ DECam bandpasses as expected. 
The power-law fits to the data in each bandpass are all consistent with the expected $(1 + z)$ law to within $2\sigma$.

Since there is a well documented stretch-luminosity relationship in Ia light curves \citep{Phillips1993, Phillips1999, Kasen2007}, it is possible that Malmquist bias could skew the data to larger widths at high redshift where we may not see the less-luminous SNe. Regardless, this does not greatly influence the quality of our fits since the DES SN data extend to such high redshifts that the intrinsic dispersion in widths is significantly smaller than the time dilation signal. In fact, we find that the standard deviation in the width residuals (i.e.\ of $w_i - (1 + z_i)^{1.003}$ for all SNe) within Fig.~\ref{fig:widthAve} is only $\sim 0.15$ (which includes observational uncertainty as well as intrinsic stretch dispersion). 
At $z\sim 1$ we would expect a time dilation factor of $\sim 2$ and so the contribution of time dilation far outweighs the intrinsic light curve stretch in the supernova population.  

\citet{Nicolas2021} \citep[see also][]{Howell2007} showed that there is a $5\sigma$ `redshift drift' in the stretch of their unbiased multi-survey sample of SNe~Ia; that is, higher redshift SNe~Ia tend to be intrinsically wider. Although we do not see this trend in the DES-SN5YR sample, we nevertheless quantify how such a drift could affect a time dilation measurement (Appendix~\ref{app:x1drift}). Given that the drift is so small, we find its impact would be minimal even if it is hidden in the data ($|\Delta b| \lesssim 0.02$). 

As a test of the robustness of our method, we reran all the width-fitting code with a requirement of pre-peak observations in each light curve. At \emph{most}, this changed the power law fit by $\Delta b = -0.004$ for the $g$ band. The calculated $b$ values in the other bands were increased by one or two thousandths (including the averaged fit of Fig.~\ref{fig:widthAve}), or not at all. Interestingly, including this pre-peak restriction reduced our number of unique SNe widths by only 24 in total. This reduction does not let us conclusively say if this method is robust at fitting light curves without pre-peak data, and future analyses may look at purposefully degrading the dataset (e.g. by manually removing pre-peak data) to investigate this.

Our method creates a unique reference light curve for each SN as a function of bandpass and redshift, and so we are able to infer a time dilation signature no matter the photometric band. This is in contrast to \citet{Goldhaber2001} who showed time dilation in the $B$-band and suggested it would hold in the other bands \citep[see also][]{Wang2003}. Fig.~\ref{fig:WidthComparison} compares the calculated widths in each band relative to the $i$-band sample (which has the most SNe of any DECam band). The apparent discrepancies toward high widths in the $i$ band might be explained by dust effects or noise domination in the observation of high redshift supernovae \citep{Moller2022MNRAS, Moller2024arXiv}. With that said, we see generally broad agreement between the widths in the different bands as expected of our source wavelength-flexible model.

{\new To avoid de-redshifting reference light curves we devised another method, similar in concept to method two above, that would yield a posterior distribution of $b$. This entailed generating a reference curve \emph{without} first scaling it in time (as in the second method), and fitting target data with a free $b$ (as opposed to $w$) to the reference data of each constituent band. That is, 
\begin{equation}
    \Delta t_\text{target} = \Delta t_\text{reference} \left(\frac{1 + z_\text{target}}{1 + z_\text{reference}}\right)^b
\end{equation}
and we attempted this with the same $\chi^2$ minimisation procedure as in equation~(\ref{eq:chisquaremin}). With this method we tried 1) separating the reference data into the constituent bands, fitting the $\chi^2$ to each band reference and minimising the weighted sum of these $\chi^2$. The sum was weighted according to how many points made up each constituent band reference curve, and this method is preferred as it assesses the fit to all bands fairly. We also tried 2) fitting $b$ using the $\chi^2$ on the entire scaled reference (i.e. all bands together), but this is not preferred as the reference is not necessarily composed of points equally sampled from all bands and so can prefer fitting $b$ towards a single bands reference (i.e. a particular redshift). This method was abandoned overall as we did not have enough data (even with the DES dataset) to accurately fit $\chi^2$ values to each band. We expect that this procedure would be viable in the future with an even larger SNe~Ia dataset of a comparable redshift distribution, or by using a Bayesian approach (which will be the topic of future work). We instead performed the analysis with flux-scatter-minimisation and width-fitting methods as the former does not rely on fitting target SN data and the latter fits to a unified (in phase) reference curve composed of all available bands.} 

Finally, in the spirit of the results of \citet{Goldhaber2001}, we similarly state that our method with the DES dataset would disfavour the null hypothesis of no cosmological time dilation to a $1.003/0.005 \simeq 200\sigma$ significance. (If $\sigma$ values were still meaningful at that extreme!)  Our uncertainty estimate is a statistical uncertainty only; in Appendix~\ref{app:x1drift} we look at a possible systematic effect due to evolution of the stretch of supernova light curves as a function of redshift. We find it to be small, with a likely upper bound of $\sigma_b^\text{sys} \simeq 0.01$. Even with an upper limit to the uncertainty of $\sigma_b^\text{sys} + \sigma_b^\text{stat} \simeq 0.015$ this remains the most precise constraint on cosmological time dilation. 

\section{Conclusions}\label{sec:conclusions}
Using two distinct methods, we have conclusively identified $(1 + z)$ cosmological time dilation using the multi-band photometry of \nSN\ SNe~Ia from the Dark Energy Survey that met our quality cuts. We make this detection with the most model-independent methods yet in the literature and with the largest survey of high redshift supernovae.

For both methods, we create a `reference curve' unique to each supernova (and each bandpass) which describes the expected light curve shape without accounting for the stretch variation associated with SN~Ia subtypes. Doing this relied on the scale of the available DES data (the number of SNe, the frequency of imaging, and the redshift range) and would not be possible with a significantly smaller survey. Creating this reference curve only relies on the assumption that SNe~Ia should be standard candles/clocks. 

Using this reference curve we show an inherent preference of $\sim(1 + z)^1$ time dilation in the data, first by minimising the flux scatter in the data via a redshift-dependent temporal scaling, and then with a more traditional light curve width estimation. The latter allows for numerical estimates with uncertainty with which we obtain a factor of $(1 + z)^{1.003\pm0.005}$ time dilation signature -- the most precise constraint on cosmological time dilation yet.  

We discuss factors and choices that affect our fits and notably see no  indication that Malmquist bias or light-curve stretch significantly impacts our results. 
Our results infer a cosmological time dilation signature aligning strongly with the expected theory, corroborating past findings \citep{Leibundgut1996, Goldhaber2001, Blondin2008, Lewis2023} with more SNe and at a higher redshift than ever before.

\section*{Contribution Statement}
RMTW performed the main analysis and drafted the manuscript; TMD supervised the project, performed the redshift drift analysis, and wrote/edited sections of the paper; GFL assisted in testing the code and developing ideas; PS, CL were internal reviewers and RK the final reader. JDV, HTD, KG, and AM gave feedback on the draft manuscript; AM, DB, RK advised on the data and analysis; the remaining authors have made contributions to this
paper that include, but are not limited to, the construction of DECam
and other aspects of collecting the data; data processing and calibration; developing broadly used methods, codes, and simulations;
running the pipelines and validation tests; and promoting the science
analysis.

\section*{Acknowledgements}
RMTW, TMD, RCa, SH, acknowledge the support of an Australian Research Council Australian Laureate Fellowship (FL180100168) funded by the Australian Government. AM is supported by the ARC Discovery Early Career Researcher Award (DECRA) project number DE230100055.

Funding for the DES Projects has been provided by the U.S. Department of Energy, the U.S. National Science Foundation, the Ministry of Science and Education of Spain, 
the Science and Technology Facilities Council of the United Kingdom, the Higher Education Funding Council for England, the National Center for Supercomputing 
Applications at the University of Illinois at Urbana-Champaign, the Kavli Institute of Cosmological Physics at the University of Chicago, 
the Center for Cosmology and Astro-Particle Physics at the Ohio State University,
the Mitchell Institute for Fundamental Physics and Astronomy at Texas A\&M University, Financiadora de Estudos e Projetos, 
Funda{\c c}{\~a}o Carlos Chagas Filho de Amparo {\`a} Pesquisa do Estado do Rio de Janeiro, Conselho Nacional de Desenvolvimento Cient{\'i}fico e Tecnol{\'o}gico and 
the Minist{\'e}rio da Ci{\^e}ncia, Tecnologia e Inova{\c c}{\~a}o, the Deutsche Forschungsgemeinschaft and the Collaborating Institutions in the Dark Energy Survey. 

The Collaborating Institutions are Argonne National Laboratory, the University of California at Santa Cruz, the University of Cambridge, Centro de Investigaciones Energ{\'e}ticas, 
Medioambientales y Tecnol{\'o}gicas-Madrid, the University of Chicago, University College London, the DES-Brazil Consortium, the University of Edinburgh, 
the Eidgen{\"o}ssische Technische Hochschule (ETH) Z{\"u}rich, 
Fermi National Accelerator Laboratory, the University of Illinois at Urbana-Champaign, the Institut de Ci{\`e}ncies de l'Espai (IEEC/CSIC), 
the Institut de F{\'i}sica d'Altes Energies, Lawrence Berkeley National Laboratory, the Ludwig-Maximilians Universit{\"a}t M{\"u}nchen and the associated Excellence Cluster Universe, 
the University of Michigan, NSF's NOIRLab, the University of Nottingham, The Ohio State University, the University of Pennsylvania, the University of Portsmouth, 
SLAC National Accelerator Laboratory, Stanford University, the University of Sussex, Texas A\&M University, and the OzDES Membership Consortium.

Based in part on observations at Cerro Tololo Inter-American Observatory at NSF's NOIRLab (NOIRLab Prop. ID 2012B-0001; PI: J. Frieman), which is managed by the Association of Universities for Research in Astronomy (AURA) under a cooperative agreement with the National Science Foundation.

The DES data management system is supported by the National Science Foundation under Grant Numbers AST-1138766 and AST-1536171.
The DES participants from Spanish institutions are partially supported by MICINN under grants ESP2017-89838, PGC2018-094773, PGC2018-102021, SEV-2016-0588, SEV-2016-0597, and MDM-2015-0509, some of which include ERDF funds from the European Union. IFAE is partially funded by the CERCA program of the Generalitat de Catalunya.
Research leading to these results has received funding from the European Research
Council under the European Union's Seventh Framework Program (FP7/2007-2013) including ERC grant agreements 240672, 291329, and 306478.
We  acknowledge support from the Brazilian Instituto Nacional de Ci\^encia
e Tecnologia (INCT) do e-Universo (CNPq grant 465376/2014-2).

This manuscript has been authored by Fermi Research Alliance, LLC under Contract No. DE-AC02-07CH11359 with the U.S. Department of Energy, Office of Science, Office of High Energy Physics.

\section*{Data Availability}

The data are available on Zenodo and GitHub as described in the DES supernova cosmology paper \citep{DES-SN5YR} and DES-SN5YR data release paper \citep{Sanchez2024}. 
The generated width fits and associated uncertainties for all \nSN SNe are included in the analysis GitHub (see Code Availability section), as are supplementary plots not included in the paper.

\section*{Code Availability}
Our code in all analysis and plotting relied on the open source Python packages \texttt{NumPy} \citep{Numpy2020}, \texttt{Matplotlib} \citep{Matplotlib2007}, \texttt{Pandas} 
\citep{Pandas2023}, and \texttt{SciPy} \citep{Scipy2020} -- specifically the Nelder-Mead algorithm described in \citet{Nelder1965}. \\
The code used to generate the width fits/reference curves and all associated figures is available at \href{https://github.com/ryanwhite1/DES-Time-Dilation}{github.com/ryanwhite1/DES-Time-Dilation}.

\bibliographystyle{mnras}
\bibliography{references}


\appendix
\section{Stretch drift with redshift}\label{app:x1drift}
\begin{figure}
    \centering
    \includegraphics[width=84mm]{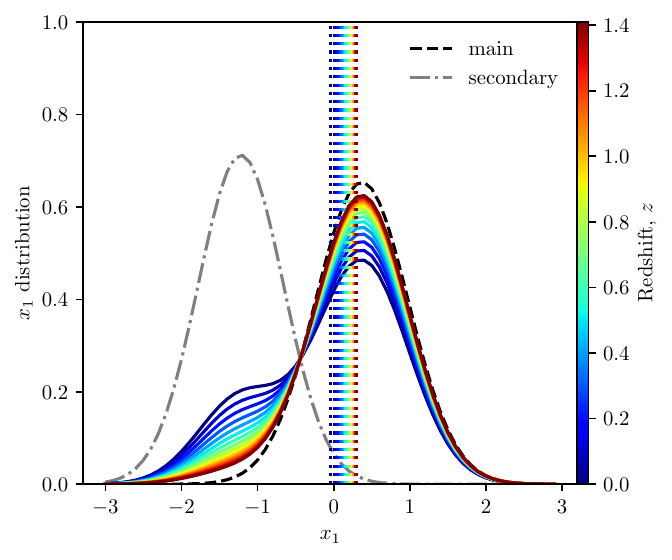}
    \includegraphics[width=84mm]{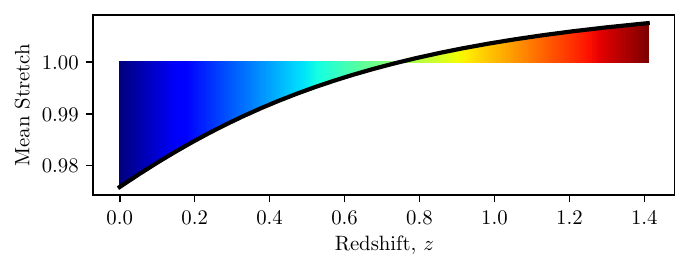}
    \caption{Upper panel: Distribution of $x_1$ values predicted by  \citet{Nicolas2021}.  The grey and black dashed Gaussians show the two components of the supernova population.  The coloured lines show the total distribution for several different redshifts.  The vertical dashed lines show the mean of the redshift distribution (in the same colours as the legend). One can see that the mean drifts from low to high $x_1$ as redshift increases.  Lower panel: The black line shows the evolution of the mean stretch ($s$) of the supernova population with redshift, where colours match the redshifts in the upper legend.  The intrinsic light curve width is proportional to $s$, and therefore light curves are expected to be about 3\% wider at $z=1$ than at $z=0$.  This is much less than the factor of two widening due to time dilation. }
    \label{fig:x1drift}
\end{figure}

There is evidence that the stretch distribution of SNe evolves with redshift, as the fraction of older and younger progenitors evolves.  \citet{Nicolas2021} give the following relation for the evolution of the SN stretch distribution,
\beq P(x_1)=\delta(z)\mathcal{N}(\mu_1,\sigma_1^2) + (1-\delta(z))\left(a\mathcal{N}(\mu_1,\sigma_1^2) + (1-a)\mathcal{N}(\mu_2,\sigma_2^2)\right), \label{eq:Px1}\eeq
where $\mathcal{N}(\mu,\sigma^2)$ is a normal distribution with mean $\mu$ and variance $\sigma^2$ and the values of the parameters were $(a,\mu_1,\mu_2,\sigma_1,\sigma_2, K)=(0.51,0.37,-1.22,0.61,0.56,0.87)$. The fraction of young supernovae in the population is given by,
\beq\delta(z)=\left(K^{-1}(1+z)^{-2.8} +1\right)^{-1}.	\label{eq:s} \eeq
where the parameter $K$ is related to the star formation rate within galaxies \citep{Rigault2013A&A}. The distribution given by equation~(\ref{eq:Px1}) is shown in the upper panel of Fig.~\ref{fig:x1drift} for several redshifts, where the vertical dashed lines show the resulting change in the mean $x_1$.  The relationship between $x_1$ and the stretch of the supernova is given by \citep{guy2007},
\beq s = 0.98 + 0.091 x_1 + 0.003 x_1^2 - 0.00075x_1^3\eeq
and this is shown in the lower panel of Fig. ~\ref{fig:x1drift}.

Since light curve width is directly proportional to stretch, that means that light curves at redshift $z=1$ should be approximately 3\% wider than those at $z=0$.  This is therefore substantially sub-dominant to the factor of 2 widening expected from time-dilation over the same range.  

Compounding this effect, supernovae with wider light curves tend to be brighter, so selection effects might also cause you to find a shift to wider light curves at higher redshifts.  

\begin{figure}
    \centering
    \includegraphics[width=84mm]{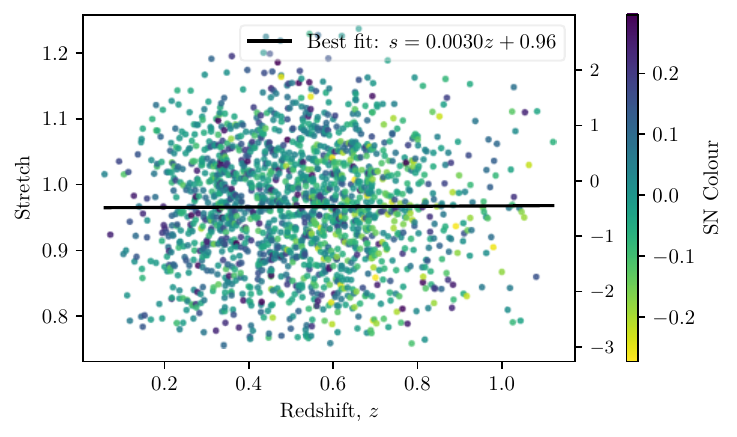}
    \caption{The distribution of stretch in the DES-SN5YR data as a function of redshift (calculated from the SALT3 fitted $x_1$ values in the SN rest frame using equation~\ref{eq:s}), with $x_1$ shown on the right axis.  Fitting a straight line to this distribution shows no significant trend in the stretch with redshift.  }
    \label{fig:x1data}
\end{figure}

Nevertheless, in the DES-SN5YR data there is no indication that the mean stretch parameter $x_1$ changes with redshift, see Fig.~\ref{fig:x1data}. The stretch values shown in this plot are from the SALT3 fits to the light curves and are in the rest frame of the supernova (i.e. corrected for time dilation).

\begin{figure}
    \centering
    \includegraphics[width=84mm]{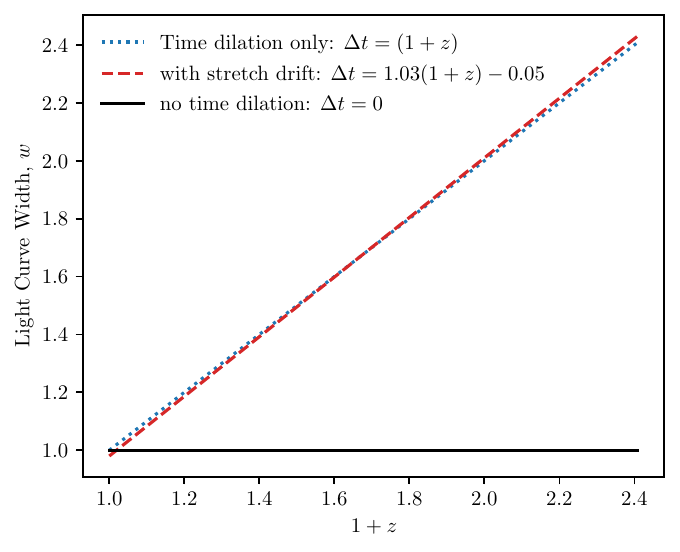}
    \caption{The effect of adding the predicted stretch evolution of SNe Ia vs redshift is to cause the width vs redshift plot to be slightly steeper.  If this result is present we therefore expect to slightly overestimate $b$, as we will attribute that widening to time dilation.  }
    \label{fig:x1drift-td}
\end{figure}

Despite the consistency of $x_1$ in the DES sample we want to quantify how large the {\em potential}  drift in the light curve widths could be if equation~(\ref{eq:Px1}) holds. Thankfully this over-estimation can be readily quantified. When you make a mock data set with this intrinsic widening included you find you would actually get a line of $\Delta t \approx 1.03(1+z) -0.05$ (see Fig.~\ref{fig:x1drift-td}).  In other words, it would change the slope by $\sim$3\%. 
This is in contrast to the recovered linear model fit in the Fig.~\ref{fig:widthAve} caption, hence indicating that this redshift-dependent stretch is not evident in the DES-SN5YR data.

The impact of high-redshift supernovae tending to have a few percent wider stretch than their low-redshift counterparts would cause us to slightly overestimate $b$.  The magnitude of the impact on $b$ depends on your redshift distribution, we estimate a shift of $|\Delta b|\lesssim 0.01$ for the DES data, and we consider this a likely upper limit to the systematic uncertainty on our result.  Since our aim in this paper is to fit the light curves with the minimal modelling assumptions (and since we do not see an $x_1$ trend in our light curve fits) we have chosen {\em not} to correct for this trend.  Instead we note that any potential effect would only be a small deviation around the slope of $w/(1+z)\sim 1$ that we see.  

\section{Reference Curve Selection Derivation} \label{app:reference}
We begin with the definition of redshift, 
\begin{equation}
    1 + z = \frac{\lambda_o}{\lambda_e} \label{eq:redshift}
\end{equation}
where $z$ is the source redshift, $\lambda_o$ is the observed wavelength of light and $\lambda_e$ is the original emission wavelength of light. If we are sampling from a band with central wavelength $\lambda_r$ during reference curve construction, we need to find a central redshift $z_r$ for our reference SNe selection, given that we are fitting the target light curve in a band of central wavelength $\lambda_f$. The idea is to match the original emitted wavelengths and so we can divide by another instance of equation~(\ref{eq:redshift}),
\begin{equation}
    \frac{1+z}{1 + z_r} = \frac{\lambda_f / \lambda_e}{\lambda_r / \lambda_e} = \frac{\lambda_f}{\lambda_r}
\end{equation}
Then, we can rearrange to find an expression for our target redshift $z_r$,
\begin{align}
    \lambda_r (1 + z) &= \lambda_f (1 + z_r) \\
    z_r &= \frac{\lambda_r(1 + z)}{\lambda_f} - 1 \label{eq:targetredshift}
\end{align}
We can then append a term $\pm \Delta z$ on equation(~\ref{eq:targetredshift}) to give us a range of applicable redshift values as in Section~\ref{sec:ref_curve}. Finally, it is useful in the broader context of the paper (and Fig.~\ref{fig:referencepop}) to show this redshift range in terms of some fraction of the band FWHM of the band that the target SN was observed in, $\delta \Delta \lambda_f$. To do this we set $\Delta z = \delta \Delta \lambda_f / \lambda_f$ and shift the term into the fraction within equation~(\ref{eq:targetredshift}),
\begin{equation}
    z_r = \frac{\lambda_r (1 + z) \pm \delta\Delta \lambda_f}{\lambda_f} - 1
\end{equation}
which yields the redshift sampling range of equation~(\ref{eq:referencerange}) that we use in the analysis.

\section{Null test --- no de-redshifting of reference light curves} 
\label{app:nonderedshifted}

\begin{figure*}
    \centering
    \includegraphics[width=0.85\textwidth]{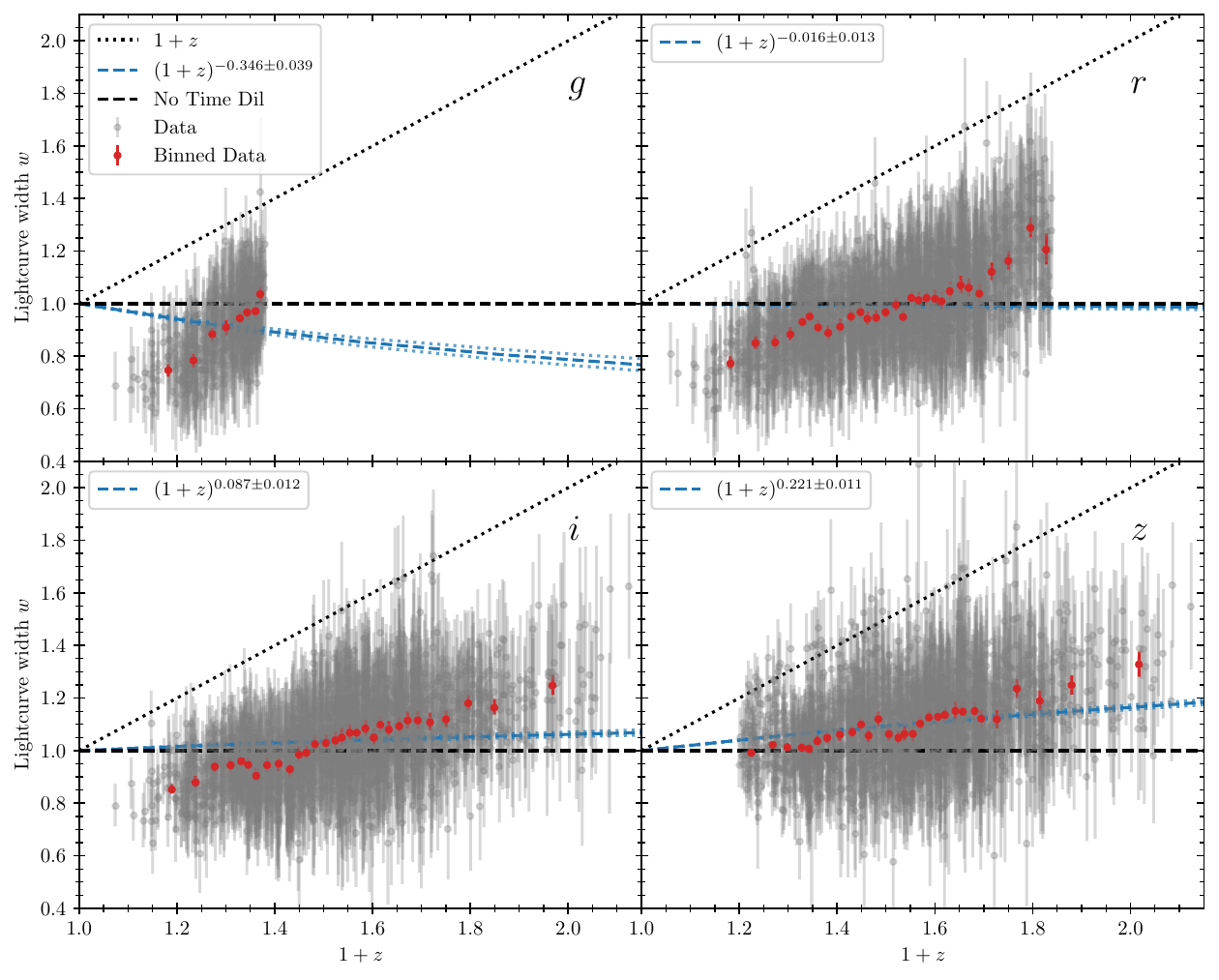}
    \caption{Light curve widths measured with respect to a reference curve that has {\em not} been de-time-dilated.  We nevertheless still see a persistent trend of increasing light curve width with redshift. The vertical offset from the $(1+z)$ line arises because the non-de-time-dilated reference curves are wider than rest-frame light curves, i.e. this offset is yet another indication of time dilation.  The black horizontal dashed line indicates no time dilation and the blue dashed lines are (poor) $(1 + z)^b$ model fits to the data.  If there was no time dilation, these fits would be horizontal lines with $b=0$. }
    \label{fig:allwidths_nonderedshifted}
\end{figure*}

{\new To confirm that our method is able to rule out no time dilation we repeated the analysis without de-redshifting the reference curves.  That means that the reference curves would look like the top panel of Fig.~\ref{fig:referenceconstruction}.  If the data were not time dilated, then we should fit consistently $b=0$ in this case.  Fig.~\ref{fig:allwidths_nonderedshifted} clearly shows that this null test fails. Because time-dilation is present in the data, it means that the reference curves are now wider than they should be --- making the measured light curve widths narrower.  Despite this, the trend for higher-redshift light curves to be wider persists, in very strong contradiction to the no-time-dilation hypothesis. }

\section*{Affiliations}
$^{1}$ School of Mathematics and Physics, University of Queensland,  Brisbane, QLD 4072, Australia\\
$^{2}$ Sydney Institute for Astronomy, School of Physics, A28, The University of Sydney, NSW 2006, Australia\\
$^{3}$ Center for Astrophysics $\vert$ Harvard \& Smithsonian, 60 Garden Street, Cambridge, MA 02138, USA\\
$^{4}$ Institut d'Estudis Espacials de Catalunya (IEEC), 08034 Barcelona, Spain\\
$^{5}$ Institute of Space Sciences (ICE, CSIC),  Campus UAB, Carrer de Can Magrans, s/n,  08193 Barcelona, Spain\\
$^{6}$ Centre for Astrophysics \& Supercomputing, Swinburne University of Technology, Victoria 3122, Australia\\
$^{7}$ Department of Physics and Astronomy, University of Pennsylvania, Philadelphia, PA 19104, USA\\
$^{8}$ Centre for Gravitational Astrophysics, College of Science, The Australian National University, ACT 2601, Australia\\
$^{9}$ The Research School of Astronomy and Astrophysics, Australian National University, ACT 2601, Australia\\
$^{10}$ Department of Physics, Duke University Durham, NC 27708, USA\\
$^{11}$ School of Physics and Astronomy, University of Southampton,  Southampton, SO17 1BJ, UK\\
$^{12}$ Department of Physics \& Astronomy, University College London, Gower Street, London, WC1E 6BT, UK\\
$^{13}$ Institute of Cosmology and Gravitation, University of Portsmouth, Portsmouth, PO1 3FX, UK\\
$^{14}$ Cerro Tololo Inter-American Observatory, NSF's National Optical-Infrared Astronomy Research Laboratory, Casilla 603, La Serena, Chile\\
$^{15}$ Laborat\'orio Interinstitucional de e-Astronomia - LIneA, Rua Gal. Jos\'e Cristino 77, Rio de Janeiro, RJ - 20921-400, Brazil\\
$^{16}$ Fermi National Accelerator Laboratory, P. O. Box 500, Batavia, IL 60510, USA\\
$^{17}$ Instituto de F\'{\i}sica Te\'orica, Universidade Estadual Paulista, S\'ao Paulo, Brazil\\
$^{18}$ Departamento de F\'isica Te\'orica and Instituto de F\'isica de Part\'iculas y del Cosmos (IPARCOS-UCM), Universidad Complutense de Madrid, 28040 Madrid, Spain\\
$^{19}$ University Observatory, Faculty of Physics, Ludwig-Maximilians-Universit\"at, Scheinerstr. 1, 81679 Munich, Germany\\
$^{20}$ Department of Astronomy and Astrophysics, University of Chicago, Chicago, IL 60637, USA\\
$^{21}$ Kavli Institute for Particle Astrophysics \& Cosmology, P. O. Box 2450, Stanford University, Stanford, CA 94305, USA\\
$^{22}$ SLAC National Accelerator Laboratory, Menlo Park, CA 94025, USA\\
$^{23}$ Instituto de Astrofisica de Canarias, E-38205 La Laguna, Tenerife, Spain\\
$^{24}$ INAF-Osservatorio Astronomico di Trieste, via G. B. Tiepolo 11, I-34143 Trieste, Italy\\
$^{25}$ Institut de F\'{\i}sica d'Altes Energies (IFAE), The Barcelona Institute of Science and Technology, Campus UAB, 08193 Bellaterra (Barcelona) Spain\\
$^{26}$ Hamburger Sternwarte, Universit\"{a}t Hamburg, Gojenbergsweg 112, 21029 Hamburg, Germany\\
$^{27}$ Centro de Investigaciones Energ\'eticas, Medioambientales y Tecnol\'ogicas (CIEMAT), Madrid, Spain\\
$^{28}$ Department of Physics, IIT Hyderabad, Kandi, Telangana 502285, India\\
$^{29}$ Jet Propulsion Laboratory, California Institute of Technology, 4800 Oak Grove Dr., Pasadena, CA 91109, USA\\
$^{30}$ Institute of Theoretical Astrophysics, University of Oslo. P.O. Box 1029 Blindern, NO-0315 Oslo, Norway\\
$^{31}$ Kavli Institute for Cosmological Physics, University of Chicago, Chicago, IL 60637, USA\\
$^{32}$ Instituto de Fisica Teorica UAM/CSIC, Universidad Autonoma de Madrid, 28049 Madrid, Spain\\
$^{33}$ Center for Astrophysical Surveys, National Center for Supercomputing Applications, 1205 West Clark St., Urbana, IL 61801, USA\\
$^{34}$ Department of Astronomy, University of Illinois at Urbana-Champaign, 1002 W. Green Street, Urbana, IL 61801, USA\\
$^{35}$ Santa Cruz Institute for Particle Physics, Santa Cruz, CA 95064, USA\\
$^{36}$ Center for Cosmology and Astro-Particle Physics, The Ohio State University, Columbus, OH 43210, USA\\
$^{37}$ Department of Physics, The Ohio State University, Columbus, OH 43210, USA\\
$^{38}$ Australian Astronomical Optics, Macquarie University, North Ryde, NSW 2113, Australia\\
$^{39}$ Lowell Observatory, 1400 Mars Hill Rd, Flagstaff, AZ 86001, USA\\
$^{40}$ Departamento de F\'isica Matem\'atica, Instituto de F\'isica, Universidade de S\~ao Paulo, CP 66318, S\~ao Paulo, SP, 05314-970, Brazil\\
$^{41}$ George P. and Cynthia Woods Mitchell Institute for Fundamental Physics and Astronomy, and Department of Physics and Astronomy, Texas A\&M University, College Station, TX 77843,  USA\\
$^{42}$ LPSC Grenoble - 53, Avenue des Martyrs 38026 Grenoble, France\\
$^{43}$ Instituci\'o Catalana de Recerca i Estudis Avan\c{c}ats, E-08010 Barcelona, Spain\\
$^{44}$ Department of Astrophysical Sciences, Princeton University, Peyton Hall, Princeton, NJ 08544, USA\\
$^{45}$ Observat\'orio Nacional, Rua Gal. Jos\'e Cristino 77, Rio de Janeiro, RJ - 20921-400, Brazil\\
$^{46}$ Department of Physics, Carnegie Mellon University, Pittsburgh, Pennsylvania 15312, USA\\
$^{47}$ Department of Physics and Astronomy, Pevensey Building, University of Sussex, Brighton, BN1 9QH, UK\\
$^{48}$ Universit\'e Grenoble Alpes, CNRS, LPSC-IN2P3, 38000 Grenoble, France\\
$^{49}$ Department of Physics, University of Michigan, Ann Arbor, MI 48109, USA\\
$^{50}$ Computer Science and Mathematics Division, Oak Ridge National Laboratory, Oak Ridge, TN 37831\\
$^{51}$ Department of Astronomy, University of California, Berkeley,  501 Campbell Hall, Berkeley, CA 94720, USA\\
$^{52}$ Lawrence Berkeley National Laboratory, 1 Cyclotron Road, Berkeley, CA 94720, USA\\
$^{53}$ ASTRAVEO LLC, PO Box 1668, MA 01931, USA\\
$^{54}$ Applied Materials Inc., 35 Dory Road, Gloucester, MA 01930, USA

\bsp	
\label{lastpage}
\end{document}